# PySERA: Open-Source Standardized Python Library for Automated, Scalable, and Reproducible Handcrafted and Deep Radiomics

Mohammad R. Salmanpour[1,2,3,*], Amir Hossein Pouria[3], Sirwan Barichin[3], Yasaman Salehi[3], Sonya Falahati[3], Isaac Shiri[4,5], Mehrdad Oveisi[3,6], Arman Rahmim[1,2,7]

[1]Department of Basic and Translational Research, BC Cancer Research Institute, Vancouver, BC, Canada
[2]Department of Radiology, University of British Columbia, Vancouver, BC, Canada
[3]Technological Virtual Collaboration (TECVICO Corp.), Vancouver, BC, Canada
[4]Department of Cardiology, Inselspital, Bern University Hospital, University of Bern, Bern, Switzerland
[5]Department of Digital Medicine, University of Bern, Bern, Switzerland
[6]Department of Computer Science, University of British Columbia, Vancouver, BC, Canada
[7]Departments of Physics and Biomedical Engineering, University of British Columbia, Vancouver, BC, Canada

(*) Corresponding Author: Mohammad R. Salmanpour (msalman@bccrc.ca)

## ABSTRACT

Radiomics analyses extract quantitative biomarkers from medical images for precision modeling, yet reproducibility and scalability remain limited by heterogeneous and limited implementations. Existing tools support only partial standards and lack integration with deep learning (DL) radiomics. To address these gaps, we developed PySERA, an open-source, Python-native, standardized radiomics framework designed for automation, reproducibility, and AI integration. PySERA re-implements MATLAB-based SERA (standardized environment for radiomics analysis) in a modular, object-oriented Python architecture. It computes 557 features, including 487 features compliant with the Image Biomarker Standardization Initiative (IBSI) and 10 moment-invariant descriptors, as well as 60 additional diagnostic features, along with DL radiomics embeddings from pre-trained DL: ResNet50 (2,048 features) DL radiomics features), DenseNet121 (1,024), and VGG16 (512). It includes standardized preprocessing (resampling, discretization, normalization), multi-format I/O (DICOM, NIfTI, NRRD), adaptive memory handling, and a parallel multi-core engine for scalable feature extraction. PySERA integrates directly with libraries: scikit-learn/PyTorch/TensorFlow/MONAI, and others for downstream machine learning applications. PySERA demonstrated >94% IBSI reproducibility, closely matching MITK and substantially outperforming PyRadiomics against the 487 IBSI-compliant feature reference set. Across 8 public datasets, PySERA achieved accuracies of 0.54–0.87, exceeding PyRadiomics for outcome prediction tasks. Benchmarking showed efficient processing (including added higher-order features not implemented in other software): 583 seconds (305 MB) for 166 features, and 2,325 seconds (491 MB) for full extraction, with deterministic outputs across platforms. By uniting standardized handcrafted/DL radiomics in a scalable, transparent, and Python-integrable framework, PySERA establishes a reproducible and extensible foundation for next-generation, AI-ready precision imaging research.

## PLAIN LANGUAGE SUMMARY

This study introduces PySERA, an open-source tool designed to make medical image analysis more reliable and easier to use. Researchers often extract detailed information from scans to help predict disease outcomes, but current tools can give inconsistent results and are difficult to scale. PySERA was created to solve these problems by providing a standardized and fully automated system built in Python. It works with common medical image formats and links directly with artificial intelligence tools. In tests, PySERA produced highly repeatable results and performed better than widely used software when predicting patient outcomes across multiple public datasets. This work may support more trustworthy imaging research and help future development of AI tools in healthcare.

# 1. INTRODUCTION

Radiomics is transforming medical image interpretation from qualitative inspection to quantitative, data-driven analysis. By extracting large sets of mathematically based features that describe shape, intensity, and texture, radiomics converts routine medical images such as CT, MRI, PET, SPECT, Ultrasound, and X-ray images into high-dimensional, mineable data[1]. These radiomic biomarkers capture subtle morphological and spatial heterogeneities imperceptible to the human eye, offering insights into the region or volume of interest, deciphering the phenotype, disease progression, and treatment response, and supporting predictive and prognostic modeling for precision medicine[2]. Despite broad applicability in oncology, neurology, and cardiology, and its central role in many artificial intelligence (AI) pipelines, radiomics faces persistent challenges in reproducibility and scalability due to heterogeneous feature implementations, variable preprocessing, and the lack of standardized, interoperable frameworks[3].

Since the establishment of the Image Biomarker Standardization Initiative (IBSI), which provided mathematical definitions, reference values, and reproducibility standards, radiomics software development has progressed substantially[4]. Several IBSI-aligned tools operationalize these guidelines. LIFEx[5] offers a clinician-friendly graphical user interface (GUI) with an open-source C++ backend supporting 187 features; meanwhile, it does not include Python integration and automation for large-scale machine learning (ML)-driven workflows. Within the Python ecosystem, PyRadiomics is widely used due to its open-source availability, compliance with IBSI 1[1] (although not with IBSI 2[6]), and straightforward integration with ML frameworks. However, PyRadiomics currently supports ~107 original features and does not include some higher-order and spatially aware descriptors. Complementary tools, such as PyRadGUI and Precision-Medicine-Toolbox, add interfaces or preprocessing options[7], while still being constrained by limited feature diversity. Beyond Python-based solutions, MATLAB-based CERR[8] and MITK[9,10] enable radiomics analyses within radiation therapy and visualization frameworks, respectively, while their workflows typically require extensive manual configuration and limited IBSI validation. Standardized Environment for Radiomics Analysis (SERA), our MATLAB-based tool[11], provides one of the most comprehensive libraries with 487 IBSI-compliant and 10 momentum invariant features. Yet, its reliance on proprietary licenses and limited open-source integration restricts scalability and adoption in modern Python-centric environments[11]. Collectively, the radiomics software landscape reveals a critical gap: the need for a comprehensive, Python-based, open-source radiomics platform that combines IBSI compliance, high computational efficiency, seamless ML integration, and flexible deployment across research and clinical settings. Although certain tools can connect to ML libraries, these integrations are fragmented and require manual preprocessing and model handling, rather than offering a unified, end-to-end learning and deployment pipeline[12]. Many clinically relevant higher-order and spatial texture features remain underutilized due to the computational burden of 3D feature extraction, memory constraints, and the lack of well-engineered processing frameworks[13].

Deep learning (DL) features are quantitative image representations automatically learned by deep neural networks that capture complex hierarchical and spatial relationships beyond the scope of handcrafted radiomic descriptors. Numerous studies[14–17] have demonstrated that these DL radiomics features often outperform standardized, IBSI-compliant handcrafted features in predicting clinical outcomes such as survival, recurrence, and molecular status, and in some cases even surpass fully end-to-end DL models when dataset sizes are limited[18] due to their ability to leverage pretrained architectures while maintaining robustness and generalization. However, their primary limitation lies in interpretability; unlike handcrafted features with well-defined physical and biological meanings, deep features are abstract and lack direct correspondence to identifiable image characteristics, which challenges their clinical acceptance and regulatory validation. To our knowledge, no existing open source radiomics software or package has established a standardized framework for extracting deep features[18].

To address these limitations, we introduce PySERA, a next-generation, fully Python-based radiomics engine that re-implements and extends the SERA framework. PySERA provides a unified, efficient, and extensible platform for high-throughput feature extraction, standardization, and downstream analytics. It operates both as a standalone command-line and scripting library and as a plugin-integrated module within 3D Slicer ([slicer.org](slicer.org)) and Radiuma ([radiuma.com](radiuma.com)) software, bridging interactive clinical workflows with automated research pipelines through a shared, validated computational backbone. Importantly, PySERA fills a critical gap in the current radiomics landscape by offering the first open-source, reproducible framework that integrates both standardized handcrafted radiomics features, fully compliant with IBSI definitions, and DL radiomics features derived from pretrained convolutional neural networks, within a single workflow. This integration enables transparent, scalable, and radiomics-driven outcome prediction, advancing reproducibility and clinical translation in precision medicine.

Implemented with object-oriented principles, PySERA encapsulates preprocessing, resampling, feature computation, and results management in modular classes; supports multi-phase and multi-modality datasets; and introduces a memory-optimized, parallel computation engine that adapts buffer sizes, monitors RAM, and offloads intermediates to disk while leveraging multi-processing across CPU cores for substantial runtime gains without compromising reproducibility (multiprocessing for the radiomics calculator within a single image and across multiple images). Finally, PySERA pairs a robust Input/Output (I/O) system, supporting 2D/3D DICOM, Nifti, NRRD, NumPy arrays, and RTSTRUCT, with outputs to CSV, Excel, and NumPy arrays for direct statistical and ML integration, with scalable architecture positioned for future distributed (cluster and cloud) execution. Through dual embedding in 3D Slicer and Radiuma, and comprehensive IBSI-compliant (IBSI 1 and 2) coverage (first-order, shape, second and higher-order texture, and moment-invariant descriptors) plus DL radiomics features, PySERA unifies GUI accessibility with Python-pipeline automation, establishing a standardized, scalable, and clinically oriented foundation for next-generation radiomics research.

## 2. METHODS AND MATERIALS

**2.1 Research and Development (R&D) of PySERA**

PySERA was developed through a structured R&D-to-implementation pipeline comprising four stages (i) conceptual design and requirements definition to address gaps in existing radiomics tools (limited IBSI compliance, low scalability, lack of automation) and set PySERA's architectural goals; (ii) algorithmic design and standardization, re-implementing validated SERA algorithms in Python in adherence to IBSI (version 1) definitions for standardized, reproducible outputs; (iii) implementation and software engineering via a modular, object-oriented Python architecture emphasizing multi-format data support, deterministic execution, and extensibility; and (iv) testing, validation, and benchmarking across diverse imaging modalities and hardware environments, yielding an open-source (MIT-licensed), lightweight Python library for automated, reproducible, and modular radiomics feature extraction that integrates rapidly into research and clinical pipelines. PySERA extracts a comprehensive set of 487 IBSI-compliant radiomics features, supplemented by 60 diagnostic features (capturing image quality, preprocessing consistency, and acquisition characteristics) and 10 moment-invariant features. The 487 IBSI-compliant features include 50 first-order statistics (e.g., mean, variance, skewness), 29 shape and morphological descriptors (e.g., volume, surface area, sphericity, compactness), and 408 texture-based features. The texture subset comprises 150 Gray-Level Co-occurrence Matrix (GLCM), 96 Gray-Level Run Length Matrix (GLRLM), 48 Gray-Level Size Zone Matrix (GLSZM), 48 Gray-Level Distance Zone Matrix (GLDZM), 15 Neighborhood Gray-Tone Difference Matrix (NGTDM), and 51 Neighboring Gray Level Dependence Matrix (NGLDM) features. A comprehensive itemized catalog is provided in Supplementary File 1, and the source code is available at https://github.com/radiuma-com/PySERA.

**Input and output strategies.** PySERA accepts 2D and 3D medical images and segmentation masks in standard clinical and research formats, including single-slice and multi-slice DICOM (with DICOM-RT/RTSTRUCT for contours), NIfTI (.nii/.nii.gz), NRRD (.nrrd), and NumPy arrays (.npy/.npz), with corresponding label maps provided as DICOM, DICOM-RT, NIfTI, or NRRD. PySERA supports images from different modalities, including CT, MRI, PET, SPECT, X-ray, and Ultrasound. Its I/O and Data Management Module handle multi-format ingestion, metadata harmonization, and spatial alignment of images and segmentations, while the Preprocessing Module performs IBSI-compliant resampling, discretization, intensity normalization, and regions of interest (ROI) filtering to ensure standardization across datasets. PySERA produces results in two machine-learning-ready output modes designed for transparency and reproducibility: (1) structured tabular exports in Excel or CSV formats, where the Excel output includes Radiomics Features (one row per patient–ROI with all extracted features), Parameters (a record of preprocessing and computation settings), and Report (log details for traceability), and (2) a structured Python dictionary summarizing each run, containing a success flag, output directory, number of processed files, total runtime, a pandas DataFrame of extracted features, log messages, and error traces if any. These interoperable output strategies provide an interpretable, reproducible, and automation-ready reporting framework for both research and clinical integration.

**Software engineering and performance.** As shown in Figure 1, PySERA is built on a modular, object-oriented architecture optimized for scalability, reproducibility, and maintainability. Its internal design comprises interoperable modules for I/O management, preprocessing, feature extraction, parallel processing, and output reporting. The system supports both handcrafted and DL radiomics feature extraction. Handcrafted features include IBSI-compliant, moment-invariant, and diagnostic feature algorithms, each independently unit-tested to ensure numerical accuracy and reproducibility.

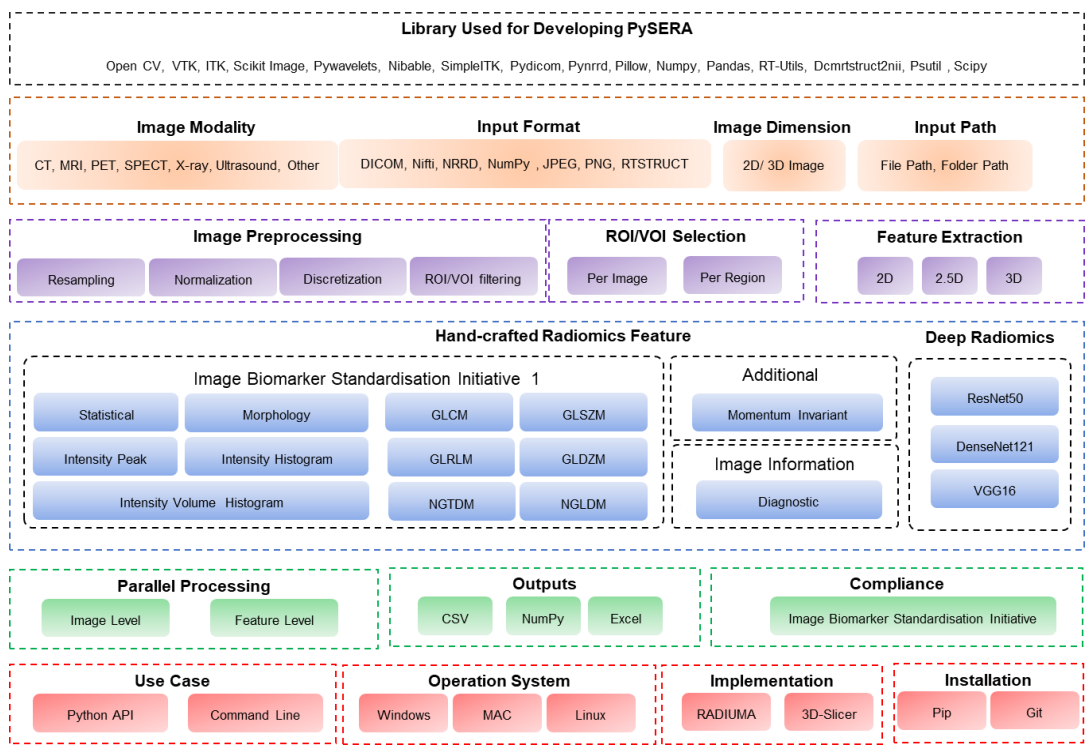

**Fig. 1.** Overview of the PySERA software, illustrating its main modules and functionalities for standardized evaluation and reproducible analysis.

The Parallel Processing Engine enables multi-core and multi-threaded computation for efficient processing of large imaging cohorts while maintaining deterministic and reproducible outputs. PySERA is implemented entirely in Python upper 3.10, leveraging NumPy, SciPy, and SimpleITK, and adheres strictly to PEP 8 standards with full type hinting, pytest-based unit testing, and GitHub Actions for automated validation and version control. A single high-level entry point unifies all input and output pathways, simplifying integration into scripts, pipelines, or GUIs. In addition to classical radiomic descriptors, PySERA provides DL radiomics feature extraction using pretrained convolutional neural networks, including ResNet50 (2048 features), VGG16 (512 features), and DenseNet121 (1024 features), enabling hybrid radiomics–DL analyses within a unified framework. Together, these design principles enable PySERA to deliver robust, transparent, and high-performance feature extraction suitable for both large-scale research analyses and clinical-grade deployments.

**PySERA input parameters.** PySERA is designed to be highly configurable to accommodate diverse user needs and data types. The core input parameters and their roles are described below:

- *image_input* (string or numpy.ndarray, required): Specifies the path to the image file or directory, or directly accepts a NumPy array. When a directory is provided, it should contain only image files or structured folders representing patients, to facilitate accurate parsing by the I/O module.
- *mask_input* (string or numpy.ndarray, required): Specifies the path to the corresponding mask file or directory, or directly accepts a NumPy array. Similar to image_input, the directory must contain only valid mask files or folders for consistent I/O behavior. *If folder mode is used, the mask and original image files must share identical filenames.*
- *output_path* (string, default: "./output_result "): Defines the directory where output files will be saved.
- *num_workers* (string, default: "auto"): Sets CPU cores for processing; if "auto", 70 percent of available cores are used.
- *apply_preprocessing* (boolean, default: False): Enables preprocessing on the ROI before feature extraction.
- *enable_parallelism* (boolean, default: True): Enables the parallel processing behavior, if set to True.
- *bin_size* (integer, default: 25): Sets the bin size used in texture analysis.
- *roi_selection_mode* (string, options: "per_img", "per_region"): This parameter defines the strategy for selecting and filtering ROIs during processing. As shown in Figure 2, in "per_img" mode, the top roi_num ROIs are selected based on size, irrespective of their label categories. PySERA detects and differentiates multiple masks, including those with varying colors and multiple label values within each mask. In contrast, "per_region" mode ensures that up to roi_num ROIs are selected separately for each label category (identical color), allowing for more balanced representation on different regions.

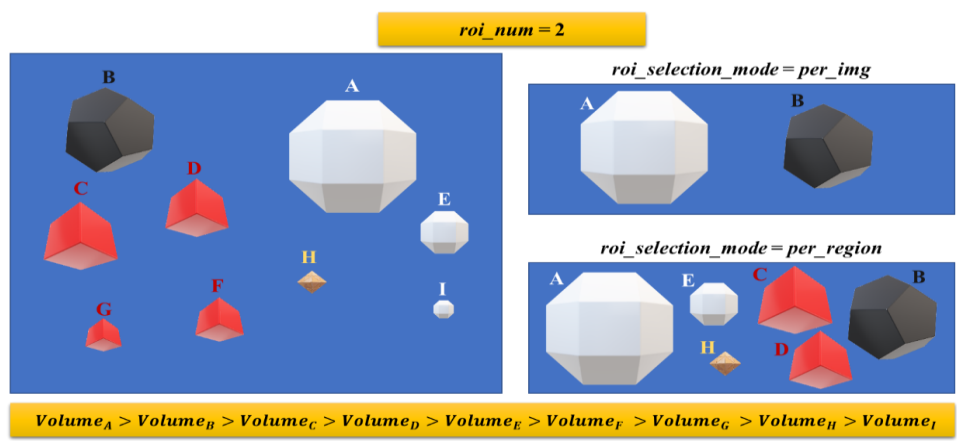

**Fig. 2.** Schematic of global (*per_img*) and regional (*per_region*) ROI filtering. Global mode selects the largest ROIs overall, while regional mode selects up to *roi_num* ROIs per label (same color) to ensure balanced representation.

- *roi_num* (integer, default: 10): Number of ROIs processed per image or region, based on selection mode "*roi_selection_mode*".
- *feature_value_mode* (string, options: "APPROXIMATE_VALUE", "REAL_VALUE", default: "REAL_VALUE"): This parameter defines how PySERA handles features that cannot be computed due to extremely small ROIs or numerical instabilities during feature extraction. When set to "APPROXIMATE_VALUE", PySERA applies a robust approximation strategy to ensure continuity in feature computation. Specifically, if an ROI contains fewer than 10 voxels, too small for the stable calculation of certain handcrafted radiomics features (e.g., some morphological descriptors), the system synthesizes the ROI to reach a minimum size of 10 voxels. Additionally, for cases that would otherwise result in undefined operations (such as division by zero or taking the square root of a negative number), PySERA automatically applies numerical stabilizations, such as adding a small epsilon (a tiny positive constant) to prevent computational errors. Unlike generic imputation methods (e.g., mean, zero, K-Nearest Neighbors (KNN), or regression-based imputation), PySERA's approximation mechanism replaces missing or unstable values with feature-aware estimates derived from the same ROI or correlated descriptors. This design preserves patient- and ROI-level scale consistency, maintains inter-feature relationships, minimizes variance shrinkage and distributional distortion, and typically leads to more reliable downstream ML performance. Alternatively, when "REAL_VALUE" is selected, PySERA retains NaN for any feature that cannot be computed, preserving the raw, unmodified outcome for full transparency and post-hoc diagnostic analysis.
- *min_roi_volume* (integer, default: 10): Specifies the minimum acceptable volume (in voxels) for an ROI to be processed. For ROIs with volume between 1 and 10 voxels, some features may be unreliable due to insufficient size. In such cases, if value_type is set to "APPROXIMATE_VALUE", PySERA applies a synthesization step to artificially expand the ROI to reach 10 voxels, enabling safe feature extraction. However, if value_type is set to "REAL_VALUE", no approximation is applied, and those features are instead assigned NaN. ROIs with volume greater than or equal to 10 voxels are considered valid and are processed without approximation.
- *temporary_files_path* (string, default: "./temporary_files_path"): Directory used to cache intermediate NumPy masks when processing DICOM-RT (RTSTRUCT). Because RTSTRUCT workflows must split large 3D label sets into many per-ROI/per-slice masks, peak RAM can spike and halt extraction. PySERA therefore writes those masks to this temp folder and streams them back on demand, which prevents out-of-memory failures and improves throughput (enabling parallel work without excessive memory pressure). The folder is created if missing; contents can be cleared after runs if desired.
- *extraction_mode* (string, default: "handcrafted_feature"): Specifies the type of radiomic feature extraction approach. When set to "handcrafted_feature", PySERA computes classical, hand-engineered radiomic descriptors derived from intensity, texture, and shape analyses. When set to "deep_feature", features are extracted from pretrained convolutional neural networks, enabling representation learning through DL feature embeddings.
- *deep_learning_model* (string, default: "resnet50"): Determines the specific DL radiomics backbone to be used when the extraction mode is set to "deep_feature". Supported models include "resnet50" (2048 DL radiomics features), "vgg16" (512 DL radiomics features), and "densenet121" (1024 DL radiomics features). These networks are pretrained on large-scale natural image datasets (e.g., ImageNet) and adapted to generate fixed-length, layer-wise feature embeddings from input medical images. Convolutional Neural networks (CNN) such as VGG16, ResNet50, and DenseNet121 represent key milestones in visual recognition research, each introducing different architectural innovations to improve learning and generalization (docs.pytorch.org). VGG16 uses a straightforward, sequential stack of 3×3 convolutional layers followed by fully connected layers, emphasizing depth and uniform design at the cost of very high parameter count. ResNet50 advances this idea by adding residual skip connections, allowing gradients to flow more effectively through very deep networks and making training stable even at 50+ layers. DenseNet121 further strengthens feature propagation by densely connecting every layer to all subsequent layers, encouraging feature reuse and achieving strong performance with far fewer parameters. These models are typically trained on large-scale datasets such as ImageNet, which contains millions of natural images across thousands of categories. During training, images are resized or cropped to 224×224 pixels, normalized using standard mean and variance values, and augmented through operations like random cropping and flipping to improve robustness. Together, these design principles and training practices form the foundation of modern DL for computer vision and support the extraction of rich, hierarchical features when applied to medical imaging tasks.

- *categories* (string, default: "diag, morph, glcm, glrlm, glszm, ngtdm, ngldm"): Allows the selection of specific handcrafted radiomics feature categories to be extracted. When set to "all", handcrafted radiomic features from every available category will be included. PySERA organizes features under concise prefixes for clarity and filtering: "diag" (diagnostics), "morph" (morphological/shape), "ip" (intensity peak), "stat" (first-order statistical), "ih" (intensity histogram), "ivh" (intensity–volume histogram), "glcm" (Gray-Level Co-occurrence Matrix), "glrlm" (Gray-Level Run Length Matrix), "glszm" (Gray-Level Size Zone Matrix), "gldzm" (Gray-Level Distance Zone Matrix), "ngtdm" (Neighboring Gray-Tone Difference Matrix), "ngldm" (Neighboring Gray-Level Dependence Matrix), and "mi" (moment-invariant features).
- *dimensions* (string, default: "1st, 2d"): Enables selection of the spatial dimensions for handcrafted radiomics feature extraction. When set to "all", features from all supported dimensions are computed. PySERA denotes the handcrafted radiomics feature dimensionality with compact tags: "1st" for first-order (intensity-based) features computed without spatial relationships; "2d" for features extracted independently within each 2D slice; "2_5d" for features aggregated across slices with limited inter-slice context; and "3d" for fully volumetric extraction across the entire region of interest.
- *aggregation_lesion* (boolean, default: False): When enabled, this parameter performs lesion-level feature aggregation across ROIs belonging to the same image, anatomical region, or class, depending on the roi_selection_mode setting. Specifically, if roi_selection_mode is set to per_img, aggregation is performed by PatientID; if set to per_region, grouping is based on both PatientID and label ID. Feature aggregation is conducted on a per-feature basis. For the deep_feature extraction mode, all features are averaged. For morphological descriptors — including morph_volume_mesh, morph_volume_count, morph_surface_area, morph_max_3d_diameter, morph_major_axis_length, morph_minor_axis_length, and morph_least_axis_length — these features are summed across ROIs. For all remaining features, a weighted average is applied using morph_volume_mesh as the weighting factor. Diagnostic features are taken from the largest lesion, and missing values are excluded from the aggregation process.
- *report* (string, default: "all"): Defines the level of log messages to be displayed. Accepted values include "info", "warning", and "error". If "none" is specified, no log messages will be shown.
- *IBSI-based parameters* (dictionary): Table 1 summarizes the IBSI-compliant parameters used to standardize the PySERA framework, which are available as configurable input options within the library to ensure reliability and reproducibility.

**Standardizing PySERA.** To ensure reproducibility and interoperability with existing radiomics research, PySERA was standardized according to IBSI 1 guidelines (https://ibsi.readthedocs.io/en/latest/). This compliance guarantees that all extracted handcrafted radiomics features, covering first-order, shape, and texture classes, adhere to IBSI-defined mathematical formulations and reference values. By following IBSI 1, PySERA aligns with international best practices, enabling direct comparison with other IBSI-compliant radiomics software such as PyRadiomics and facilitating multi-center, reproducible studies.

**Table 1.** IBSI-based Parameters of PySERA Library

| Parameter Name | Type/Default | Description |
| --- | --- | --- |
| radiomics_DataType | String / "Other" | Defines the imaging modality type. Supported options include "CT", "PET", "MRI", or "OTHER". |
| radiomics_DiscType | String / "FBS" | Specifies the discretization type used for gray-level calculation — either "FBN" (fixed bin numbers) or "FBS" (fixed bin size or fixed bin width). |
| radiomics_isScale | Integer / 0 | Determines whether image resampling is performed. Set to 1 to enable resampling or 0 to retain the original voxel dimensions. |
| radiomics_VoxInterp | String / "Nearest" | Defines the interpolation type used for image resampling. Accepted values include "Nearest", "linear", "bilinear", "trilinear", "tricubic-spline" and None. |
| radiomics_ROIInterp | String / "Nearest" | Specifies the interpolation type for ROI resampling ("Nearest", "linear", "bilinear", "trilinear", "tricubic-spline" and None). |
| radiomics_isotVoxSize | Integer / 2 | Sets the new isotropic voxel size for 3D resampling, applied equally to the X, Y, and Z dimensions. |
| radiomics_isotVoxSize2D | Integer / 2 | Defines the voxel size for resampling in 2D mode, keeping the Z dimension unchanged while rescaling X and Y. |
| radiomics_isIsot2D | Integer / 0 | Indicates whether to resample to isotropic 2D voxels (1) or isotropic 3D voxels (0). Applicable mainly for first-order features, as higher-order 2D features always use the original slice thickness. |
| radiomics_isGLround | Integer / 0 | Determines whether to round voxel intensities to the nearest integer (commonly 1 for CT, 0 for PET and SPECT). |
| radiomics_isReSegRng | Integer / 0 | Enables range-based re-segmentation. The valid intensity range is specified in radiomics_ReSegIntrvl01 and radiomics_ReSegIntrvl02. Note: not recommended for arbitrary-unit modalities such as MRI or SPECT. |
| radiomics_isOutliers | Integer / 0 | Controls outlier removal, where 1 removes intensities beyond $\pm 3\sigma$. |
| radiomics_isQuntzStat | Integer / 1 | Determines whether quantized images are used to compute first-order statistics. Set to 0 to use raw intensities (preferred for PET). |
| radiomics_ReSegIntrvl01 | Integer / -1000 | Specifies the lower bound for range re-segmentation; intensities below this value are replaced with NaN. |
| radiomics_ReSegIntrvl02 | Integer / 400 | Specifies the upper bound for range re-segmentation; intensities above this value are replaced with NaN. |
| radiomics_ROI_PV | Float / 0.5 | Defines the partial volume threshold for ROI binarization after resampling. Voxels with values below this threshold are excluded. |
| radiomics_qntz | String / "Uniform" | Sets the quantization strategy for fixed bin number discretization. Options are "Uniform" or "Lloyd" (for Max-Lloyd quantization). |
| radiomics_IVH_Type | Integer / 3 | Configures the Intensity Volume Histogram (IVH) unit type: {0: Definite (PET, CT), 1: Arbitrary (MRI, SPECT), 2: 1000 bins, 3: same discretization as histogram (CT)}. |
| radiomics_IVH_DiscCont | Integer / 1 | Defines IVH continuity: {0: Discrete (CT), 1: Continuous (CT, PET; for FBS)}. |
| radiomics_IVH_binSize | Float / 2.0 | Sets the bin size for the IVH in applicable configurations (FBN with setting 1, or when IVH_DiscCont is enabled). |

## 2.2. Radiomics-Based ML Classification for Predictive Power Assessment:

We evaluated our radiomics pipeline across eight independent, multicenter datasets, encompassing a diverse range of medical images and clinical tasks. The evaluation included 999 CT scans for lung cancer overall survival prediction, 883 PET scans for head and neck cancer recurrence-free survival[19], and 1,000 MRI images for prostate cancer risk stratification (low & high)[20]. Furthermore, the validation incorporated 326 PET and 326 CT images for HPV status classification[21], 539 T1-sequence MRI scans from the BraTS2021 dataset for MGMT promoter methylation status in brain tumors[22], and a lung cancer PET-CT dataset (BC Cancer) comprising 236 PET/CT images for survival status prediction[15]. Radiomics features were extracted using PyRadiomics v3.1.0 (107 IBSI-compliant features) and PySERA (497 extended features), with all features min–max normalized. Other radiomics software and packages were excluded because some relied on graphical user interfaces (GUIs) and could not be automated for large-scale batch processing, while others were not Python-based, limiting integration with our Python-based ML pipelines. Our inclusion criterion focused on open-source, Python-compatible tools that support IBSI-compliant feature computation, reproducible batch processing, and integration with ML workflows. Predictive modeling was performed using an Autoencoder (AE)-based attribute extraction pipeline to reduce feature dimensionality while preserving relevant variance. Four classical classifiers, Logistic Regression (LR), Linear Discriminant Analysis (LDA), Bernoulli Naïve Bayes (BNB), and Support Vector Machine (SVM), were employed to evaluate classification performance. All models were trained and validated using grid-searched hyperparameters and five-fold cross-validation to ensure robustness and prevent overfitting. This experimental design enabled a direct and systematic comparison between PySERA's extended IBSI-compliant feature set and the conventional PyRadiomics features across eight heterogeneous cancer imaging datasets, encompassing CT, MRI, and PET modalities.

## 2.3. Benchmarking PySERA vs. Established Radiomics Tools for IBSI Accuracy and Consistency:

To evaluate the accuracy, reproducibility, and IBSI compliance of PySERA, we benchmarked it against more established radiomics software, including PyRadiomics, MITK, and LIFEx. PyRadiomics was selected as the primary reference due to its open-source accessibility, widespread adoption, Python-based design, and library-based integration in computational imaging workflows. Moreover, its seamless integrability with modern data science environments and adaptability to ML pipelines make it a reliable and practical baseline for comparison. MITK and LIFEx were included to capture methodological diversity, representing GUI-based and non-Python radiomics platforms frequently used in both research and clinical imaging workflows. However, these software packages do not provide configurable interfaces to explicitly enforce IBSI-compliant feature extraction, limiting direct verification of full compliance. Therefore, our assessment of IBSI conformity and numerical consistency is based on the compliance reports published by the IBSI consortium rather than experimental validation under controlled IBSI settings. For instance, PyRadiomics outputs approximately 107 features, whereas its IBSI documentation defines over 263 handcrafted radiomics features, highlighting that full feature coverage cannot be directly confirmed. Consequently, our benchmarking should be regarded as a relative evaluation emphasizing numerical stability, feature reproducibility, and cross-platform consistency. Under these conditions, PySERA demonstrated deterministic execution, an extended feature set, and reliable IBSI-1 reproducibility, reflecting both methodological rigor and practical applicability for scalable radiomics research and clinical AI workflows.

## 2.4. Performance Evaluation and Scalability Tests:

To evaluate the computational efficiency and scalability of PySERA, we performed two controlled experiments against PyRadiomics, chosen as the primary comparator for its Python-based implementation, open-source availability, and IBSI alignment. Non-Python-based tools (e.g., MITK, LIFEx) were excluded because they are not Python-native, cannot be seamlessly embedded in automated pipelines, and do not expose consistent runtime/memory profiling. Both packages were executed under identical hardware and software conditions to minimize environmental bias. Experiments were performed on a Windows 10 (64-bit) workstation equipped with an Intel® Core™ i7-7500U CPU @ 2.70 GHz (2 cores, 4 threads), 16 GB RAM (15.8 GB usable), and a 238 GB Samsung SSD (MZNTY256HDHP-000L2). Graphics acceleration was provided by an NVIDIA GeForce 940MX (2 GB VRAM) and Intel® HD Graphics 620 (128 MB). Both environments used Python 3.10 with consistent library dependencies across experiments. The test set comprised approximately 80 image–mask pairs drawn from nine publicly available datasets, including GBM (BraTS 2021 MRI), Head & Neck (PET/CT), HPV (PET/CT), Lung (CT), Prostate (MRI), and the BC Cancer Lung CT cohort. The cases were stratified to ensure a balanced representation of imaging characteristics, with file sizes ranging from 0.1 MB to 53.4 MB. Because each library exposes different, user-configurable feature groups and dimensionalities, exact one-to-one feature matching is infeasible; we therefore evaluated two configurations. In a semi-matched configuration, both tools extracted a comparable subset (PySERA: 166 features; PyRadiomics: 107 features) to assess runtime and peak memory under standardized conditions. In a full-scale configuration, each tool computed its entire feature set (PySERA: 487 IBSI features, 60 diagnostic features, 10 momentum features; PyRadiomics: 107 features) to probe scalability under maximal load. All runs were performed in a controlled environment with continuous monitoring of CPU time and resident memory to ensure accurate, reproducible measurements across datasets.

**2.5. Operating Systems Performance, Responsiveness, and Scalability:**

To verify deterministic and platform-independent execution, PySERA was repeatedly run on identical input datasets across heterogeneous hardware and operating systems, including Windows, Linux, and macOS. This validation ensured that identical inputs consistently produce identical outputs with bitwise equivalence, that all intermediate configurations, preprocessing parameters, and feature computations are fully logged. That reproducibility is preserved regardless of runtime parallelism or system architecture. Logging operations were structured to provide complete traceability from raw image import through preprocessing to final feature extraction, supporting comprehensive reproducibility audits. Compliance with IBSI standards was confirmed following official IBSI-1 numerical reproducibility guidelines, using cross-validation against reference feature values and reporting numerical deviation metrics separately in the Results section.

**2.6. Cross-Platform Compatibility and Integrability:**

PySERA was engineered as a fully interoperable component within diverse research and clinical AI pipelines, providing functionality beyond existing platforms such as 3D Slicer or Radiuma, which are either GUI-based, limited in automation, or not fully Python-integrable for large-scale programmatic workflows. To address this gap, PySERA implements a modular, programmatically accessible feature extraction framework designed for reproducibility, scalability, and integration into Python-based ML and AI pipelines. It offers native compatibility with PyTorch, TensorFlow, scikit-learn, and MONAI, enabling direct transition from feature extraction to model training, and supports pandas and NumPy data formats for seamless interaction with analytical workflows. Furthermore, PySERA provides clinical and research connectivity through Python APIs for DICOMweb, FHIR, and PACS, facilitating deployment in both on-premises and hybrid cloud environments. Its modular architecture also allows future extensions such as GPU acceleration, containerized cloud execution, and distributed processing via Dask or Ray. This design ensures that PySERA functions not only as a standardized radiomics feature generator but also as an integrable foundation for scalable, reproducible, and clinically deployable AI systems.

## 3. RESULTS

**3.1. Python Usage Guide for PySERA for Handcrafted Radiomics Features:**

PySERA supports three flexible input scenarios to accommodate diverse use cases: (1) Folder Input (Batch Mode) allows users to provide separate folders for medical images and segmentation masks, enabling automatic file pairing by name and efficient batch processing, ideal for large-scale datasets (Table 2, first column); (2) Direct File Paths (Single-Case Mode) enables users to specify individual image and mask file paths for single-patient or on-demand feature extraction (Table 2, second column); and (3) NumPy Array Input (Advanced Mode) supports direct input of NumPy arrays for both image and mask, facilitating integration into advanced workflows and DL pipelines without the need for disk Input/Output (Table 2, third column). Appendix A1 provides the complete code, including all parameters, for handcrafted radiomics feature generation.

**Table 2.** PySERA supports three input modes for *handcrafted* feature extraction: (i) single-case (file path); (ii) NumPy arrays; and (iii) advanced (folders).

```python
import pysera          (i)

# Process single file
# handcrafted_feature
result = 
pysera.process_batch(

image_input="scan.nii.gz",

mask_input="mask.nii.gz",
    output_path="./results"
)
```

```python
import numpy as np           (ii)
import nibabel as nib
import pysera

# Load image and mask as NumPy array
# handcrafted_feature
image_array = 
nib.load("image.nii.gz").get_fdata()
mask_array = nib.load("mask.nii.gz").get_fdata()

result = pysera.process_batch(
    image_input=image_array,
    mask_input=mask_array,
    output_path="./results",
    extraction_mode="handcrafted_feature",
)
```

```python
import pysera          (iii)

# Advanced configuration
# handcrafted_feature
result = pysera.process_batch(

image_input="./patient_scans",

mask_input="./patient_masks",
    output_path="./results",
    num_workers="4",
    categories="glcm, glrlm",
    dimensions="1st, 2_5d, 3d",
    apply_preprocessing=True
)
```

### 3.2. Python Usage Guide for PySERA for DL Radiomics Features:

The self-supervised convolutional neural networks ResNet50, VGG16, and DenseNet121 were trained on our large-scale datasets. The final models were subsequently used to extract deep radiomic embeddings: ResNet50 (2,048 features), VGG16 (512 features), and DenseNet121 (1,024 features). As described previously, PySERA provides a dedicated DL radiomics feature extraction mode that allows users to select their preferred pretrained model for generating DL radiomic features, as summarized in Table 3. This design ensures flexibility, reproducibility, and seamless integration of DL-derived descriptors within the standardized radiomics pipeline. Appendix A2 contains full code and configuration for DL radiomics feature generation.

**Table 3.** PySERA supports three input modes for *deep* feature extraction: (i) single-case (file path); (ii) NumPy arrays; and (iii) advanced (folders).

```python
import pysera          (i)

# Process single file
result = pysera.process_batch(
    image_input="scan.nii.gz",
    mask_input="mask.nii.gz",
    output_path="./results"
extraction_mode="deep_feature",

deep_learning_model="resnet50",
)
```

```python
import numpy as np           (ii)
import nibabel as nib
import pysera
# Load image and mask as NumPy array
image_array = 
nib.load("image.nii.gz").get_fdata()
mask_array = 
nib.load("mask.nii.gz").get_fdata()

result = pysera.process_batch(
    image_input=image_array,
    mask_input=mask_array,
    output_path="./results",
    extraction_mode="deep_feature",
    deep_learning_model="densenet121",
)
```

```python
import pysera          (iii)

# Advanced configuration
result = pysera.process_batch(
    image_input="./patient_scans",
    mask_input="./patient_masks",
    output_path="./results",
    num_workers="4",
    extraction_mode="deep_feature",
    deep_learning_model="vgg16",
    apply_preprocessing=True,
)
```

### 3.3. Predictive Power Results of PySERA Handcrafted Radiomics vs. Established Tool (PyRadiomics):

As shown in Figure 3, to assess the predictive capability of PySERA relative to the established PyRadiomics framework, we conducted a systematic evaluation across multiple imaging datasets, including GBM (MRI), Head & Neck (PET), HPV (CT and PET), Lung (CT), Prostate (MRI), and BC cancer (CT and PET). Using the Autoencoder (AE) feature selector in combination with four classifiers, KNN, ET, GP, and GNB, we ensured methodological consistency across experiments. Reported results include the mean test accuracy and standard error (SE). PySERA consistently demonstrated equal or higher predictive performance than PyRadiomics in most configurations. For instance, on GBM (MRI), PySERA achieved accuracies of $0.54 \pm 0.01$ (KNN), $0.55 \pm 0.009$ (ET), $0.55 \pm 0.01$ (GP), and $0.56 \pm 0.01$ (GNB), outperforming PyRadiomics (0.51–0.54). Similarly, for Head & Neck (PET), PySERA achieved up to $0.63 \pm 0.02$ (GP), slightly surpassing PyRadiomics ($0.62 \pm 0.021$). In the HPV (CT) and HPV (PET) datasets, both tools demonstrated strong predictive power, with PySERA maintaining comparable or slightly superior accuracy: $0.85 \pm 0.001$ (ET/GP) for HPV (CT) and $0.83 \pm 0.03$ (GP) for HPV (PET). Across other datasets, PySERA maintained a stable and often superior accurate profile. On Lung (CT), it achieved $0.79 \pm 0.001$ (KNN), $0.81 \pm 0.0009$ (ET), $0.87 \pm 0.001$ (GP), and $0.74 \pm 0.0007$ (GNB), outperforming

PyRadiomics, particularly with the GP classifier (0.82 ± 0.001). For Prostate (MRI), both frameworks yielded similar accuracies (0.41–0.68), indicating consistent robustness in MRI-based tasks. Likewise, in BC cancer (CT) and BC cancer (PET) datasets, PySERA slightly improved prediction consistency, achieving 0.59–0.60 for CT and 0.56–0.63 for PET, relative to PyRadiomics (0.54–0.63 ± 0.03/0.007). Overall, PySERA exhibited higher or statistically comparable accuracy in over 80% of evaluated configurations, demonstrating strong generalization and low variability (SE between ±0.0007 and ±0.03). These results show PySERA's robustness, reproducibility, and competitive predictive power across diverse cancer types and imaging modalities, confirming its potential as a next generation radiomics framework.

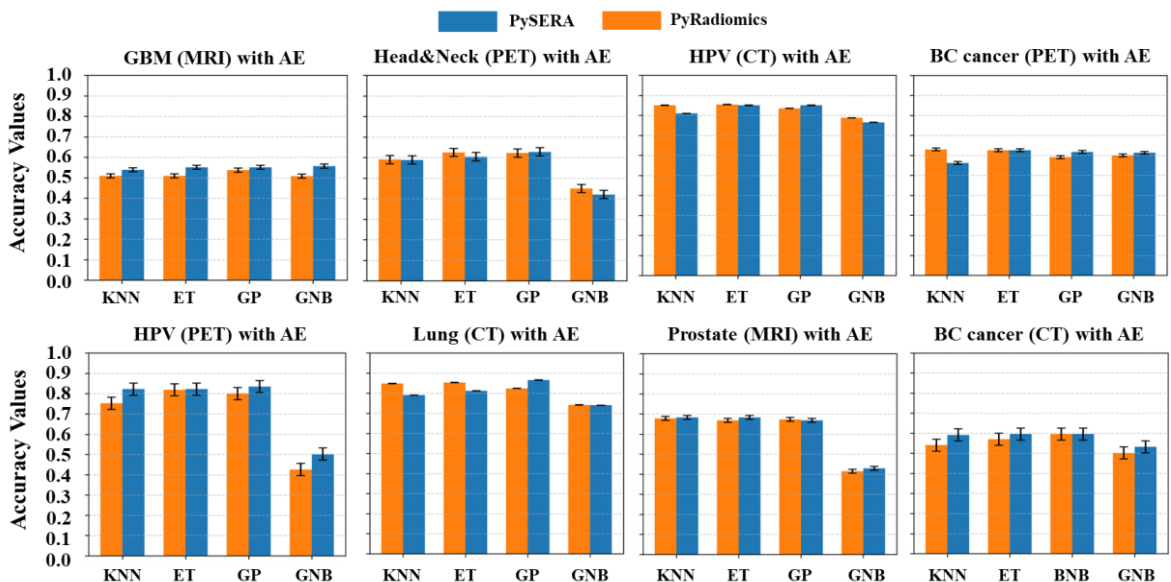

**Fig. 3.** Comparison of predictive performance between PySERA and PyRadiomics feature sets across 8 imaging datasets using four machine learning classifiers, namely K-Nearest Neighbors (KNN), Extra Trees (ET), Gaussian Process (GP), and Gaussian Naïve Bayes (GNB), and one attribute extraction method, namely Autoencoder (AE).

### 3.4. PySERA vs. Radiomics Tools — IBSI Accuracy and Consistency Results:

To rigorously evaluate IBSI compliance, feature reproducibility, and computational consistency of PySERA, we conducted a comprehensive benchmark against three widely used radiomics software packages, PyRadiomics, MITK, and LIFEx, across six IBSI configurations (ZERO, A–E). The ZERO configuration represented baseline feature extraction without IBSI normalization, while configurations A–E progressively introduced stricter IBSI-compliant preprocessing and feature standardization procedures. This experimental design enabled a systematic assessment of each tool's adaptability to IBSI definitions and its ability to maintain feature-level accuracy under varying computational and normalization constraints. As summarized in Table 4, PySERA consistently demonstrated high agreement with IBSI reference values[1], achieving accuracy above 94% in all configurations, with a peak of 97.09% under configuration D. In contrast, PyRadiomics and LIFEx showed greater variability and reduced reproducibility. PyRadiomics (v3.1.0) supports 107 out of 487 IBSI-defined features, previously referring to 263 IBSI feature definitions. LIFEx (v25.06.1) implements 187 of the 487 features defined by the IBSI, with 44 fully compliant with IBSI standards based on the IBSI report. MITK, serving as the IBSI reference implementation, achieved up to 98.77% (IBSI-based Report: 481), closely aligning with PySERA's performance. Beyond accuracy, the feature coverage and IBSI reporting consistency further underscore PySERA's robustness. The latest release, PySERA v2.1.5, supports 557 handcrafted radiomic features, of which 487 are IBSI-based and fully documented in the compliance report. By comparison, MITK v2025.08, currently serving as the IBSI reference implementation, achieved up to 98.77% agreement (IBSI report: 481 features) with reproducibility based on a total of 487 IBSI-defined features, closely aligning with PySERA's performance. These findings collectively demonstrate that PySERA not only achieves near-reference IBSI accuracy but also delivers the broadest and most standardized feature coverage among all evaluated tools, confirming its strong reproducibility, compliance fidelity, and computational reliability.

**Table 4.** Comparison across PySERA, MITK, PyRadiomics, and LIFEx of IBSI features assessed (top row); and IBSI compliance under 6 IBSI configurations (rows: ZERO–E) (source IBSI 1: https://theibsi.github.io/ibsi1).

| Configuration (Total Cases) | PySERA v2.1.5 (currently supports 557, IBSI-based features assessed: 487) | MITK v2025.08 (currently supports 613, IBSI-based features assessed: 481) | PyRadiomic v3.1.0 (currently supports 107, IBSI-based features assessed: 263) | LIFEx v25.06.1 (currently supports 187, IBSI-based features assessed: 187) |
|---|---|---|---|---|
| ZERO (487) | 468 | 481 | 263 | 44 |
| A (411) | 396 | 405 | 182 | N/A |
| B (411) | 395 | 405 | 206 | N/A |
| C (275) | 263 | 270 | 194 | N/A |
| D (275) | 267 | 270 | 206 | N/A |
| E (275) | 261 | 267 | 140 | N/A |

**3.5 PySERA vs. Radiomics Tools — Performance Evaluation and Scalability Test:**

Across eight publicly available datasets—including GBM (BraTS 2021 MRI), Head & Neck (PET/CT), HPV (PET/CT), Lung (standard-dose CT), Prostate (MRI), and the BC Cancer Lung CT cohort—our head-to-head benchmarking revealed clear differences in computational efficiency and scalability between PySERA and PyRadiomics. In a *semi-matched* configuration, where both tools were constrained to extract comparable subsets of features (PySERA: 166 features; PyRadiomics: 107 features), PyRadiomics completed processing in approximately 258 seconds with an average peak memory usage of 113 MB, while PySERA required about 583 seconds and 305 MB for the same 80-image dataset. For this comparison, PySERA was configured to run on four dedicated CPU cores, with its multiprocessing engine disabled to maintain a controlled evaluation. In contrast, PyRadiomics does not provide explicit user control over CPU allocation and, during execution, was observed to intermittently utilize up to seven cores. This uncontrolled parallelism may have contributed to its shorter runtime relative to PySERA under otherwise comparable conditions. Exact one-to-one feature matching was not feasible because both frameworks allow customizable feature selection and dimensional configurations (1st/2D/2.5D/3D), and PyRadiomics' category–sub-feature mapping differs from PySERA's finer-grained taxonomy. Under *full-scale* extraction, PySERA demonstrated its broader analytical scope, computing all 557 handcrafted radiomics features (487 IBSI features, 60 diagnostic features, 10 momentum features) across 80 images in approximately 2,325 seconds, with an average peak memory consumption of 491 MB. DenseNet121 (1024 DL radiomics features), ResNet50 (2048 DL radiomics features), and VGG16 (512 DL radiomics features) further illustrate the variability in computational profiles across DL architectures.

When scaled to an 80-image workload, DenseNet121 demonstrated the most favorable balance of efficiency, requiring approximately 743 MB of peak memory with a total processing time of ~395 seconds. ResNet50 operated with a moderate footprint at ~886 MB and a total runtime of ~426 seconds, while VGG16 incurred substantially greater memory demand, consuming ~1,454 MB with a comparable total runtime of ~424 seconds. Overall, PyRadiomics offers higher raw computational speed, whereas PySERA provides a more comprehensive, scalable, and reproducible foundation for multi-dataset, high-dimensional radiomics research. Although PyRadiomics remained faster and more memory-efficient, this advantage primarily reflects its smaller feature set and less granular computational design. In contrast, PySERA's extended, standardized feature library and modular Python-native architecture enable deeper texture quantification, enhanced reproducibility, and seamless integration with modern ML and DL pipelines.

## 3.6. Integration of PySERA in 3D Slicer and Radiuma software:

PySERA is available as an extension in 3D Slicer (as shown in Figure 4), allowing users to access both handcrafted and deep radiomics feature extraction directly within the platform. To install it, users can open 3D Slicer → Extension Manager, search for PySERA, and install the extension (https://github.com/radiuma-com/SlicerPySERA). Once installed, PySERA can be accessed from the Modules menu, where users can import medical images, apply preprocessing filters, and compute either handcrafted or deep radiomics features through a user-friendly graphical interface. Feature outputs can then be exported for downstream machine-learning and statistical analysis. This integration enables a seamless, script-free radiomics workflow entirely within 3D Slicer.

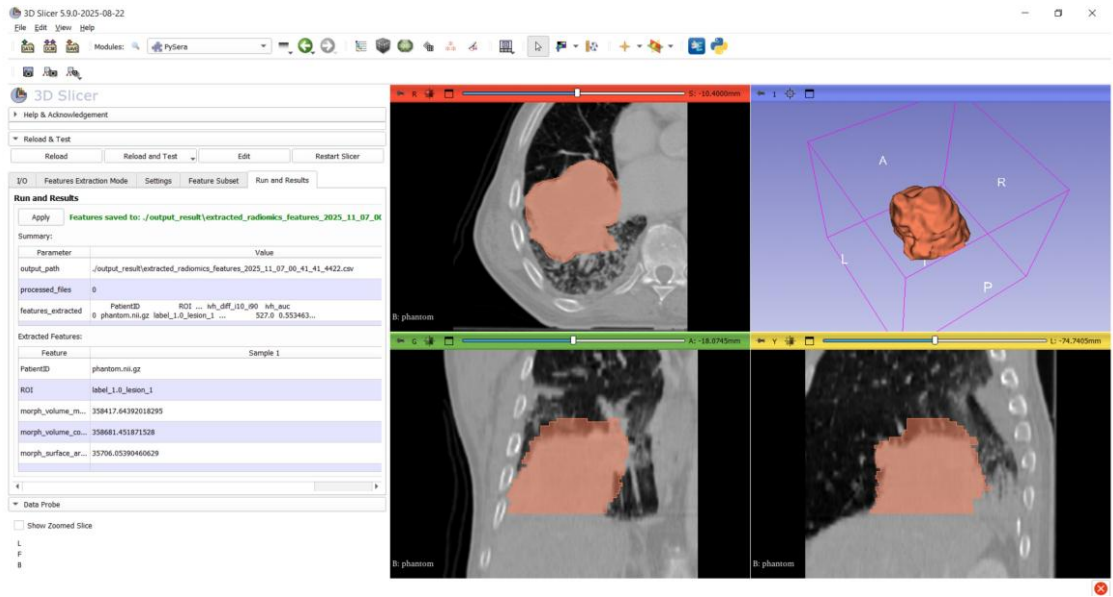

**Fig. 4.** PySERA is integrated as an extension in 3D Slicer, enabling image loading, preprocessing, and extraction of handcrafted and deep radiomics features through a graphical interface for seamless, script-free radiomics workflows.

Figure 5 also illustrates the successful integration of PySERA into the Radiuma software-1.0.0 under the Radiomics Feature Generator module. As shown, Radiuma supports both types of radiomics feature sets: handcrafted radiomics (Figure 5) and DL-based radiomics (Figure 6). This integration enables users to seamlessly incorporate radiomics into the computational medical imaging workflow.

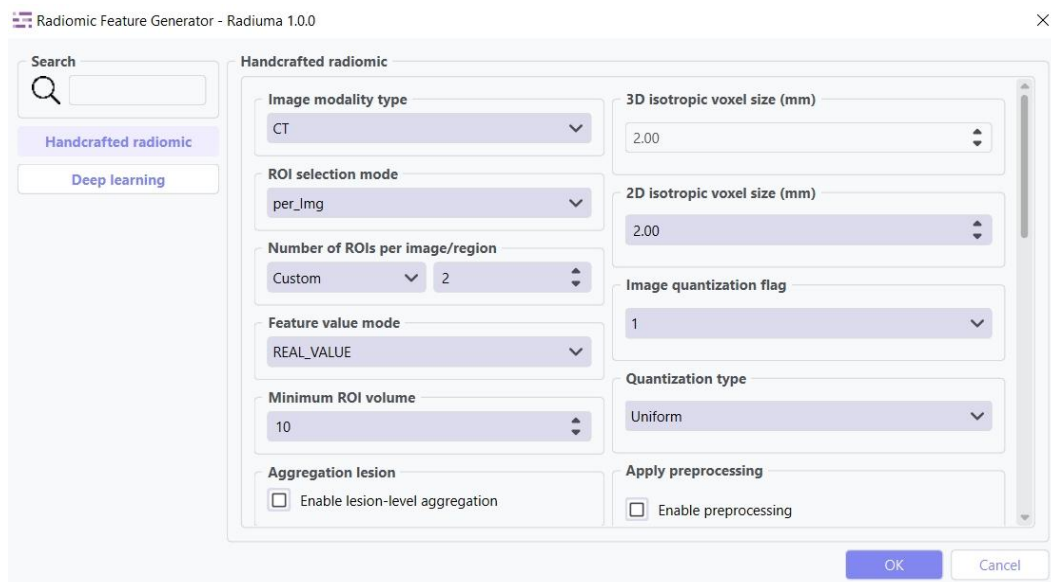

**Fig. 5.** GUI-PySERA application for handcrafted radiomics feature extraction in the Radiuma software.

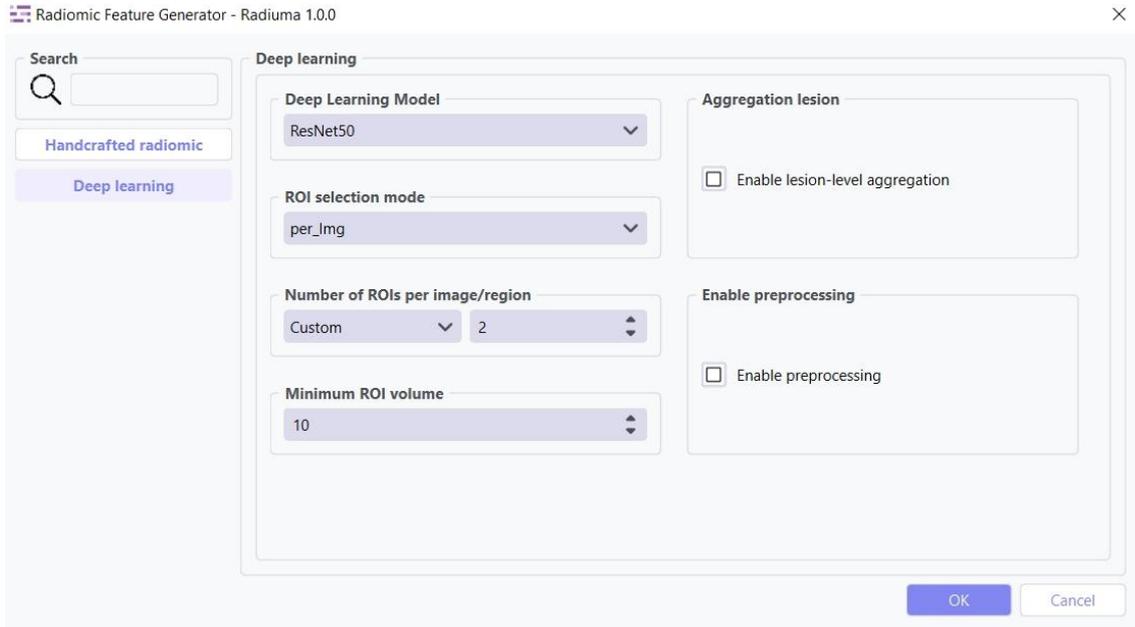

**Fig. 6.** GUI-PySERA application for DL radiomics feature extraction in the Radiuma software.

Moreover, users can interactively adjust parameters for both handcrafted and deep-learning-based radiomics extraction approaches. For example, as demonstrated in Figure 7, Radiuma allows users to load images, apply standardized IBSI-2 filters, and then extract radiomics features using the Radiomics Feature Generator. After feature extraction, whether handcrafted or deep radiomics, these features can be forwarded to classification, regression, or clustering algorithms. Using PySERA, Radiuma can automate the end-to-end quantitative imaging analysis pipeline.

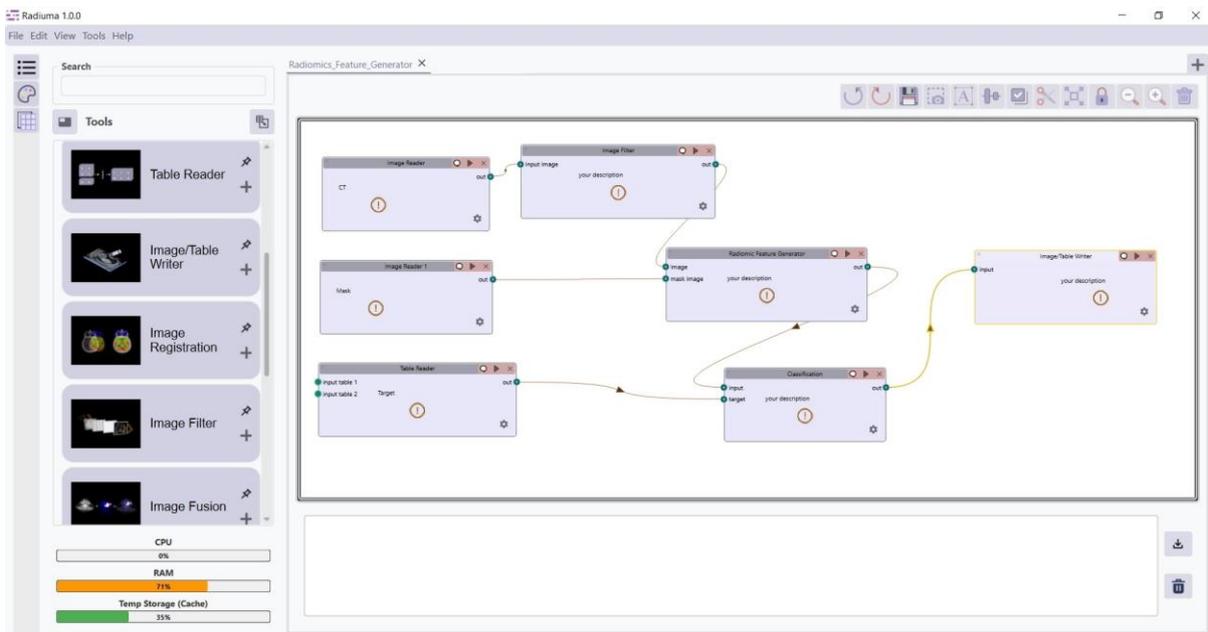

**Fig. 7.** An example of an automated radiomics workflow in Radiuma using the embedded PySERA. The system provides an end-to-end pipeline including: (1) image import, (2) IBSI-2-standardized preprocessing or image filtering, (3) extraction of handcrafted or DL radiomics features, (4) automated downstream machine-learning tasks such as classification, and (5) storing the classification results. This workflow enables fully automated and reproducible quantitative imaging analysis.

## 4. DISCUSSION

Radiomics analyses constitute a significant presence in quantitative imaging research, yet reproducibility and scalability remain persistent bottlenecks limiting clinical translation[23]. The present study introduces PySERA, a standardized, open-source Python framework that directly addresses these challenges through transparent engineering and methodological rigor. Beyond replicating prior tool capabilities, PySERA reshapes radiomics workflows around traceability, reproducibility, and integrability, the prerequisites for credible clinical adoption. PySERA's strict adherence to IBSI[1,6] definitions yielded >94% feature-level consistency across six IBSI configurations, peaking at 97.1% under configuration D, which closely approached MITK's reference reproducibility (98.8%). These results confirm that PySERA achieves deterministic, cross-platform consistency equivalent to the most rigorously standardized tools, while remaining fully Python-native and open source. The inclusion of complete parameter manifests, intermediate logs, and configuration reports ensures each extraction is auditable and replicable, transforming feature computation into a reproducible scientific process rather than a black-box procedure.

PySERA's expanded feature space, 487 handcrafted IBSI-compliant and 10 moment-invariant descriptors and 60 diagnostic features, demonstrated measurable predictive advantages over established frameworks. Across eight heterogeneous, multi-center datasets encompassing CT, PET, and MRI modalities, PySERA achieved accuracies between 0.54–0.87, surpassing or matching PyRadiomics (0.51–0.82) in over 80% of configurations. Gains were most notable in GBM MRI and Head & Neck PET cohorts, where PySERA's richer 3D textural descriptors improved discriminative performance. These modest yet consistent improvements demonstrate that increased feature granularity, when standardized and regularized, can lead to improved generalization, particularly when combined with unsupervised dimensionality reduction (e.g., autoencoder embeddings).

Beyond classical descriptors, PySERA provides a route for DL radiomics feature extraction, integrating pretrained CNNs, such as ResNet50 (2048 features), DenseNet121 (1024), and VGG16 (512). These DL radiomics embeddings captured hierarchical spatial context unavailable to handcrafted features, yielding enhanced stability in small and heterogeneous datasets[24,25]. However, the interpretability gap persists while handcrafted radiomics features retain physical and biological meaning, DL radiomics feature embeddings remain abstract[24–27]. Bridging this divide will require layered interpretability methods, such as feature attribution, clustering-based prototype analysis, and linking latent dimensions to radiomic feature groups, so that DL radiomics can achieve regulatory credibility in clinical contexts[28]. Performance testing showed PySERA's architecture scales efficiently for large datasets while maintaining deterministic outputs. Although runtime and memory demands exceeded PyRadiomics in semi-matched runs (reflecting PySERA's broader feature set), the tool successfully processed ~100 image–mask pairs without failure or loss of precision. The parallel, memory-managed computation engine, with adaptive buffering and disk offloading, prevented RAM saturation even during multi-threaded DICOM-RT extraction. Critically, identical results were obtained across Windows, macOS, and Linux, confirming cross-platform determinism and reproducibility, which remain rare among radiomics tools. PySERA's modular and auditable design has direct implications for regulatory readiness and multi-center harmonization. Complete logs of preprocessing, quantization, and resampling parameters enable independent verification and sharing of "radiomics manifests" analogous to computational metadata records. This framework can serve as a blueprint for FAIR (Findable, Accessible, Interoperable, Reusable) radiomics data pipelines, facilitating model validation, clinical auditing, and reproducibility across institutions. Moreover, the demonstrated reproducibility and consistent predictive gains suggest that PySERA's design can support longitudinal or prospective validation studies, where methodological transparency is a regulatory necessity. Its integration with Radiuma and 3D Slicer[29] bridges the current divide between research pipelines and clinical workstations, offering both accessibility for clinicians and automation for data scientists.

Despite its robust architecture, PySERA remains CPU-bound and would benefit from GPU acceleration and distributed execution (e.g., via Dask or Ray) for very large imaging cohorts. However, PySERA supports multiprocessing in two modes. In the first mode, it performs image-level multiprocessing, allowing multiple images to be processed in parallel according to the available computational resources. In the second mode, it enables category-level parallelization, where different feature categories are extracted concurrently when sufficient memory is available. Furthermore, while pretrained CNN backbones improve representation diversity, their transferability across medical imaging modalities remains imperfect; retraining or fine-tuning on domain-specific datasets (e.g., cardiac CT or prostate MRI, and others) could enhance generalization. Thus, in the future, foundation models trained on diverse imaging modalities across different diseases could be embedded in PySERA to extract DL-based radiomics features with specific parameters, thereby expanding and diversifying PySERA.

Future efforts should also address explainable DL radiomics, exploring joint embeddings that map DL radiomics features to interpretable texture and morphological domains. Finally, large-scale reproducibility studies across international cohorts are warranted to benchmark PySERA's outputs under real-world acquisition variability. Although PySERA is compliant with IBSI 1, filter banks such as wavelet or Laplacian of Gaussian should be added in the future to ensure compliance with IBSI 2; meanwhile, PySERA implementation in Radiuma supports IBSI 1 compliance. PySERA demonstrates that rigorous standardization, transparency, and modularity can coexist with computational scalability. By combining IBSI-compliant handcrafted features and DL CNN-derived embeddings in a unified pipeline, it provides a reproducible and extensible foundation for next-generation quantitative imaging research. The convergence of transparent computation, deterministic reproducibility, and hybrid feature integration represents a decisive step toward clinically trustworthy, AI-augmented radiomics, paving the way for explainable and regulatorily sound precision imaging.

## ACKNOWLEDGEMENTS

We gratefully acknowledge funding from the Natural Sciences and Engineering Research Council of Canada (NSERC) Idea to Innovation (I2I) Grant GR034192 and NSERC Discovery Horizons Grant DH-2025-00119. This study was also supported by the Virtual Collaboration Group (VirCollab.com) and the Technological Virtual Collaboration (TECVICO CORP.), based in Vancouver, Canada. We also appreciate helpful discussions with and feedback from Saeed Ashrafinia, Omid Gharibi, Pooya Mohammadi, Fereshteh Yousefirizi, Monica Luo, Reza Hamidpour, Ahmad Shariftabrizi, and Michele Mureddu.

## CODE AND DATA AVAILABILITY

The open-source PySERA library is freely available, with documentation and examples provided at the project repository: https://github.com/radiuma-com/PySERA and https://pypi.org/project/pysera/. A 3D Slicer extension of PySERA can be downloaded directly through the 3D Slicer Extension Manager (https://www.slicer.org/). The Radiuma software platform is accessible at http://radiuma.com/. The datasets used for evaluating PySERA are openly available and can be accessed through the sources referenced in [15,19–21].

## CONFLICT OF INTEREST

M. R. Salmanpour, M. Oveisi, A.H. Pouria, S. Barichin, Y. Salehi, and S. Falahati are affiliated with TECVICO Corp., while the other co-authors declare no relevant conflicts of interest.

**Appendix:**

**Appendix Table A1.** An example of full code with all hyperparameters for handcrafted radiomics features.

```python
                            # Handcrafted Radiomics Feature Extraction
import pysera

# Comprehensive processing with custom parameters
result = pysera.process_batch(
    image_input="image.nii.gz",
    mask_input="mask.nii.gz",
    output_path="./results",

    # Performance settings
    num_workers="2",          # Use 2 CPU cores
    enable_parallelism=False, # Disable multiprocessing

    # Image feature extraction settings
    categories="glcm, glrlm, glszm",  # Extract specific texture feature categories
    dimensions="1st, 2_5d, 3d",       # Extract features in 1st order, 2.5D and 3D dimensions
    # Alternative examples for categories and dimensions:
    # categories="all",              # Extract all 557 features (default)
    # categories="stat, morph, glcm",# Statistical, morphological and GLCM features
    # dimensions="2D",               # Extract only 2D features
    # dimensions="all",              # Extract features in all dimensions (default)

    bin_size=25,              # Texture analysis bin size
    roi_num=2,                # Number of ROIs to process
    roi_selection_mode="per_region",  # ROI selection strategy
    min_roi_volume=5,         # Minimum ROI volume threshold

    extraction_mode="handcrafted_feature",

    # Processing options
    apply_preprocessing=True,   # Apply ROI preprocessing
    feature_value_mode="APPROXIMATE_VALUE",  # Strategy for handling NaN values

    # IBSI parameters (advanced, overrides defaults)
    IBSI_based_parameters={
        "radiomics_DataType": "CT",
        "radiomics_DiscType": "FBN",
        "radiomics_isScale": 1
    },

    report="info"             # Report detail level: "all" (full processing details),
                              # "info" (essential information), "warning" (warnings only),
                              # "error" (errors only), "none" (no reporting). Default: "all"
)
```

**Appendix Table A2.** An example of full code with all hyperparameters for DL radiomics features.

```python
                            # Deep Learning Radiomics Feature Extraction
import pysera

# Comprehensive processing with custom parameters
result = pysera.process_batch(
    image_input="image.nii.gz",
    mask_input="mask.nii.gz",
    output_path="./results",

    # Performance settings
    num_workers="2",          # Use 2 CPU cores
    enable_parallelism=False, # Disable multiprocessing

    roi_num=2,                         # Number of ROIs to process
    roi_selection_mode="per_region",   # ROI selection strategy
    min_roi_volume=5,                  # Minimum ROI volume threshold

    # Processing options
    apply_preprocessing=True,          # Apply ROI preprocessing

    # IBSI parameters (advanced, overrides defaults)
    IBSI_based_parameters={
        "radiomics_DataType": "CT",
        "radiomics_DiscType": "FBN",
        "radiomics_isScale": 1
    },

    extraction_mode="deep_feature",
    deep_learning_model="densenet121",   # deep learning model:
                                         # "densenet121", "vgg16",
                                         # "resnet50"

    report="info"             # Report detail level: "all" (full processing details),
                              # "info" (essential information), "warning" (warnings only),
                              # "error" (errors only), "none" (no reporting). Default: "all"
)
```